\documentclass[12pt]{article}
\pdfoutput=1
\usepackage[margin=1in]{geometry}
\usepackage{mathptmx}
\usepackage{titlesec}
\usepackage{etoolbox}
\usepackage{setspace}
\usepackage{graphicx}
\usepackage[
backend=biber,
style=nature,
citestyle=numeric-comp,
sorting=none
]{biblatex}

\begin{document}
\begin{center}
{\fontsize{24}{20} \selectfont Turbulence explains the accelerations of an eagle in natural flight}

\vspace{20pt}

Kasey M Laurent$\mathrm{^{a}}$, Bob Fogg$\mathrm{^{b}}$, Tobias Ginsburg$\mathrm{^{a}}$, Casey Halverson$\mathrm{^{b}}$, Michael Lanzone$\mathrm{^{b}}$, Tricia A Miller$\mathrm{^{c,d}}$, David W Winkler$\mathrm{^{c,e}}$, Gregory P Bewley$\mathrm{^{a}}$

$\mathrm{^a}${\it Sibley School of Mechanical and Aerospace Engineering, Cornell University, Ithaca, NY 14853}

$\mathrm{^b}${\it Cellular Tracking Technologies, Rio Grande, NJ 08242}

$\mathrm{^c}${\it Conservation Science Global, West Cape May, NJ 08204}

$\mathrm{^d}${\it Department of Forestry and Natural Resources, West Virginia University, Morgantown, WV 26506}

$\mathrm{^e}${\it SABER Consulting, Monterey, CA 93942}

\vskip\bigskipamount 
\vskip\medskipamount
\leaders\vrule width \textwidth\vskip0.4pt 
\vskip\bigskipamount 
\vskip\medskipamount
\nointerlineskip

\end{center}
{\fontsize{16}{14}{\bf Abstract}}

\noindent Turbulent winds and gusts fluctuate on a wide range of timescales 
from milliseconds to minutes and longer, 
a range that overlaps the timescales of avian flight behavior, 
yet the importance of turbulence to avian behavior is unclear. 
By combining wind speed data 
with the measured accelerations of a golden eagle (\textsl{Aquila chrysaetos}) 
flying in the wild, 
we find evidence in favor of a linear relationship between the eagle's accelerations and atmospheric turbulence
for timescales between about 1/2 and 10\,s. 
These timescales are comparable to those of typical eagle behaviors, 
corresponding to between about 1 and 25 wingbeats, 
and to those of turbulent gusts both larger than the eagle’s wingspan 
and smaller than large-scale atmospheric phenomena such as convection cells. 
The eagle’s accelerations exhibit power spectra 
and intermittent activity characteristic of turbulence, 
and increase in proportion to the turbulence intensity. 
Intermittency results in accelerations that are occasionally several times stronger than gravity, 
which the eagle works against to stay aloft.
These imprints of turbulence on the bird’s movements need to be further explored to understand the energetics of birds and other volant lifeforms, 
to improve our own methods of flying through ceaselessly turbulent environments, 
and to engage airborne wildlife as distributed 
probes of the changing conditions in the atmosphere. 

\vskip\bigskipamount 
\vskip\medskipamount
\leaders\vrule width \textwidth\vskip0.4pt 
\vskip\bigskipamount 
\vskip\medskipamount
\nointerlineskip

\section*{Introduction}
Birds play, dance, tumble, soar, migrate, defend territories, hunt, and court mates, 
all while in flight \cite{Newton2007,Gerrard1989}. 
Their habitat is the atmospheric boundary layer, 
and they have adapted to its normally turbulent condition \cite{Diehl2013,Stull1988}. 
In part due to the turbulence, 
a wide range of timescales 
characterize the movements of the air through which birds fly \cite{Kaimal1994}. 
The sub-range of these timescales associated with turbulence 
poses a particular challenge 
since it is within this sub-range that 
turbulence competes with the many complex behaviors of volant lifeforms 
\cite{Richardson1978,Kerlinger1989,Shepard2013,Vansteelant2017,Poessel2018}. 
Flapping and switching between behaviors, 
for instance, take place within seconds \cite{Kress2015}. 

The relevance of turbulence to volant lifeforms is not entirely known 
\cite{Reynolds2014,Quinn2019}. 
A handful of specific observations suggest an important interaction, 
such as for turkey vultures flying through turbulence over a forest canopy \cite{Mallon2016},
for migrating songbirds encountering weather \cite{Bowlin2008}, 
and for laughing gulls in wind-tunnel experiments \cite{Tucker1972}. 
Even in this small number of cases, the results are inconsistent, 
indicating increased \cite{Tucker1972,Bowlin2008}
or decreased \cite{Mallon2016} 
flight costs attributed to turbulence. 

Relative to small-scale turbulence, 
large-scale atmospheric flows have well-known effects on flight. 
Examples include 
orographic updrafts produced by topography \cite{Nourani2017,Murgatroyd2018,Sage2019}, 
thermal updrafts \cite{Hedenstrom1993,Akos2010,Reddy2018}, 
internal waves \cite{Teets2002}, 
and fronts, 
which all tend to subsidize flight. 
These interactions are easier to observe 
in part due to the fact that the flows are slowly evolving 
and in an approximate steady-state 
relative to the time it takes for a bird or aircraft to fly through them. 

The structure of a flow leaves its mark 
on the trajectories of birds 
just as it does
on the trajectories of droplets and particles 
carried by turbulent flows \cite{Wilson1996,Toschi2009}. 
These trajectories and their time derivatives 
contain evidence of flight subsidy by birds \cite{Katzner2015} 
and of the mechanisms of particle transport \cite{Voth2002,Good2014}. 
Though particles are much simpler than birds, 
much more is known about their interactions with turbulence, 
and we therefore use them as benchmarks 
by which to evaluate an eagle's motions in flight. 

\begin{figure}[t]
\centering
\includegraphics[width=.8\linewidth]{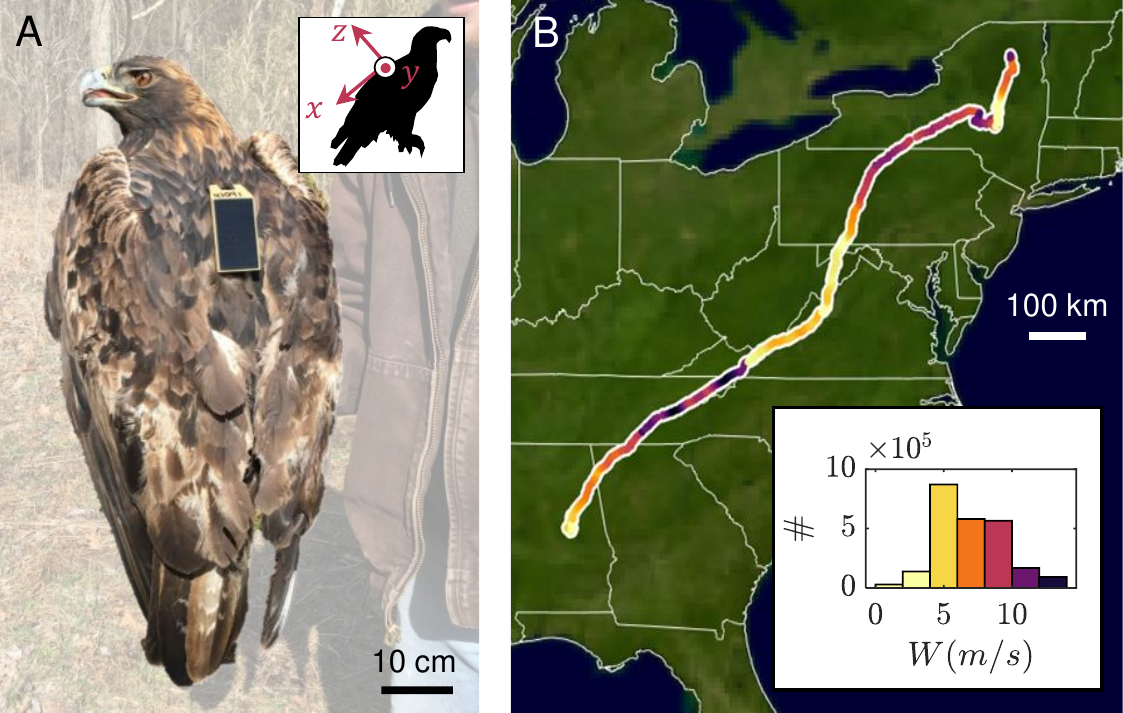}
\caption{The golden eagle, its telemetry unit and its flight path. 
(\textsl{A}) The telemetry unit mounted on the eagle's back 
collected GPS positions and tri-axial accelerations of the golden eagle 
over the course of 17 days 
(Materials \& Methods). 
\textsl{Inset:} The coordinate system ($x$, $y$, $z$) 
is fixed with respect to the telemetry unit. 
(\textsl{B}) After a winter residency in Alabama, 
the eagle migrated northeast toward Canada, 
as seen in the record of the eagle’s GPS coordinates, 
colored according to the local mean wind speed 
as given in the inset. 
Original satellite image from NASA Scientific Visualization Studio.
\textsl{Inset:} Histogram of local wind speeds extracted from a weather database 
given the eagle's positions. 
}\label{fig:bird}
\end{figure}

Turbulence is difficult to predict, sensitive to small changes in conditions, 
and evades simple description. 
Turbulence does, however, exhibit unique signatures 
including a certain distribution of energy among different scales of motion 
in combination with a lack of exact scale-invariance called intermittency \cite{Pope2000,Sreenivasan1997,Sreenivasan1999}. 
These unique features can be seen in the trajectories of 
particles carried by turbulent flows \cite{Fung2003,Mordant2004,Ayyalasomayajula2006,Calzavarini2009},
and we find analogous signatures imprinted in eagle flight trajectories. 

We analyzed observations of an adult female golden eagle ({\it Aquila chrysaetos}, Fig.~\ref{fig:bird}{\it A}) with a mass of 5\,kg and a wingspan of approximately 2\,m in natural flight from Alabama and along the Appalachian Mountains to New York from March 15 to 31, 2016, as illustrated Fig.~\ref{fig:bird}{\it B}. 
The eagle carried instruments that recorded its position and tri-axial acceleration and relayed this information 
via the mobile phone network. The eagle flew through different wind conditions, which we determined on a coarse length-scale from weather history databases given the eagle’s positions. 
Details on the estimates of wind conditions are described in the Materials \& Methods. The eagle did not move exactly with the wind for several reasons: it was large compared to some of the structures in the flow, massive relative to the air itself, and it experienced drag, produced thrust, and made decisions about which way to fly. 

\begin{figure}[t]
\centering
\includegraphics[width=8.7cm]{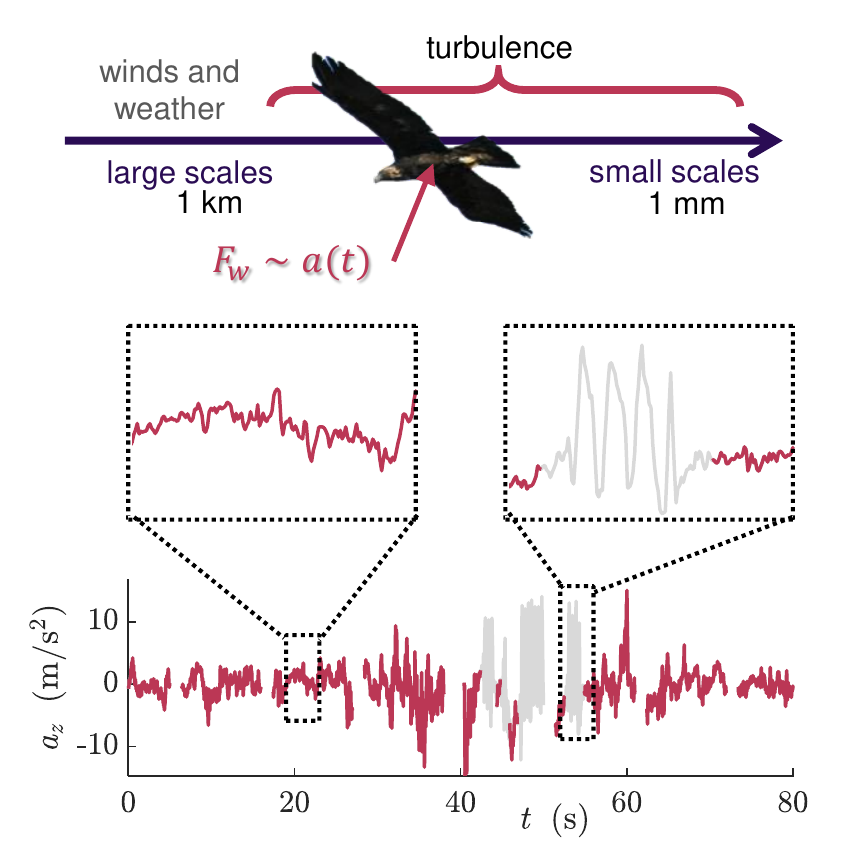}
\caption{{\it Above:} 
The eagle experiences aerodynamic forces, $F_w$, 
that fluctuate in time as it flies through the wide range of scales 
that characterize turbulence in the atmosphere. 
These forces cause accelerations, $a(t)$, recorded by the telemetry unit on the eagle. 
{\it Below:}
We explain the irregular fluctuations in the soaring accelerations (red) by the eagle's interactions with atmospheric turbulence. 
There are gaps in the data because the accelerations were recorded in bursts. 
Flapping (grey) is evident as periodic spikes and automatically separated from soaring flight 
(Materials \& Methods).
}\label{fig:flappingID}
\end{figure}

Patterns in the eagle’s accelerations and positions indicate different behaviors, such as taking-off and landing, which we classified by adapting known methods \cite{Williams2015}. 
We separated those parts of the trajectory in Fig.~\ref{fig:bird}{\it B} where the eagle was soaring from those where it was flapping its wings, which we identified as a regular oscillation in the acceleration with a frequency of about 2.8\,Hz (Fig.~\ref{fig:flappingID}). We also separated its residency in Alabama from its northward migration along the mountains. Our conclusions, however, are not sensitive to how we classified the eagle’s flight behavior.

\section*{Results}

\begin{figure}[t]
\centering
\includegraphics[width=8.7cm]{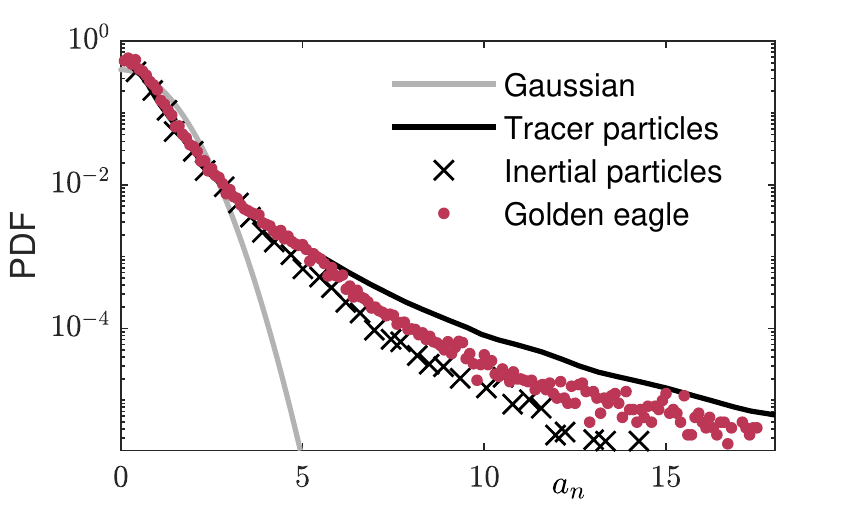}
\caption{The probability density function (PDF) of the eagle’s accelerations
while soaring is far from Gaussian 
and exhibits a long tail similar to the one for tracer particles in turbulence (–)\,\cite{Sreenivasan1997}. 
Despite their relatively high probability, 
extreme accelerations are less probable for the eagle than for tracer particles in a way consistent with the suppression of extreme events by weakly inertial particles ($\times$
, $St = 0.09 \pm 0.03$
)\,\cite{Ayyalasomayajula2006,Mordant2004}.
The acceleration distributions, $P(a_i)$, for the $x$, $y$, and $z$ directions are 
averaged, so that 
$\mathrm{P}(a_n) \equiv (1/3)(\mathrm{P}(\vert a_x / a^\prime_x \vert) + \mathrm{P}(\vert a_y / a^\prime_y \vert) + \mathrm{P}(\vert a_z / a^\prime_z \vert)$), 
where $a^\prime_i$ is the standard deviation of $a_i$. 
}
\label{fig:PDF}
\end{figure}

The soaring accelerations of the eagle were highly intermittent, as indicated by the long tails in the distribution of accelerations (Fig.~\ref{fig:PDF}). 
The way that the distributions are strongly non-Gaussian is consistent with what we know about particle accelerations in vigorous turbulence, despite the differences in scale and geometry between the particles and eagles. Tracer particles of turbulence, which follow turbulent flow exactly, show extreme accelerations many orders of magnitude more probable than a Gaussian distribution predicts. 
Increasing particle size and mass tends to depress the tails of the distribution 
in a way measured by the Stokes number ($St$), 
which is $<1$ for lightweight and small particles 
\cite{Ayyalasomayajula2006,Calzavarini2009}. 

The acceleration distribution for the eagle lies between that of tracer particles (no inertia) and that of weakly inertial particles ($St = 0.09 \pm 0.03$), 
a notion we explore further in the discussion section. 
Turbulence distributions often resemble stretched exponential functions, 
and these functions also describe the tail of the eagle’s acceleration distributions with a stretching exponent of about 1.8, 
consistent with values reported for small-scale quantities \cite{Kailasnath1992,Mordant2004,LaPorta2001}. 
The standard deviations of the $x$, $y$ and $z$ components of the accelerations 
are 0.90, 0.88 and 1.62\,$\mathrm{m/s^2}$, respectively. 

\begin{figure}
\centering
\includegraphics[width=8.7cm]{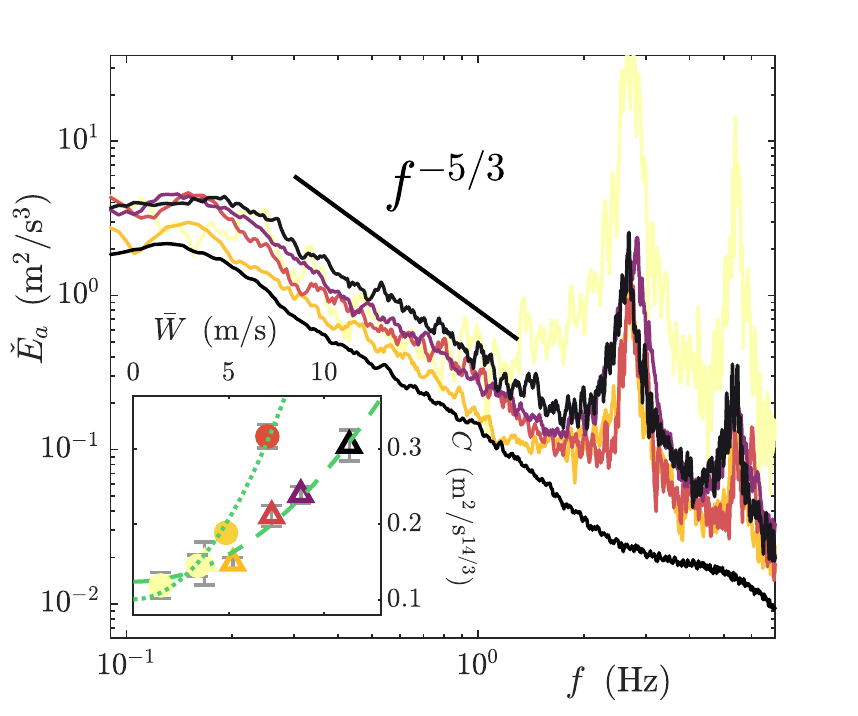}
\caption{The spectrum of the eagle’s accelerations 
scales predictably and depends on local wind speeds. 
Over an intermediate range of frequencies, 
the spectra consistently exhibit logarithmic slopes of close to -5/3rds. 
The black curve is the spectrum for all wind speeds and for only those time intervals during which the eagle was soaring and not flapping; 
flapping contributes peaks at about 2.8~Hz and its harmonics. 
The colored curves are spectra conditioned on intervals in the average wind speed, $\overline{W}$, 
in the vicinity of the eagle during migration; 
the spectrum lifts up in higher winds. 
\textsl{Inset:}
The increase in the amplitude of the spectra, $C$, 
is consistent with 
the quadratic function of the wind speed predicted by turbulence theory. 
The differences between the migrating (data: $\Delta$, model: - -) 
and winter resident (data: $\bullet$, model: $\cdot \cdot$) flight response
are discussed in Results. 
Error bars are 95\% confidence intervals. 
}
\label{fig:spectra}
\end{figure}

A dominant and reproducible feature of the acceleration spectra during migration 
and residency (Fig.~\ref{fig:spectra})
is an approximate power law that prevails between frequencies of about 0.2 and 2\,Hz. The region that follows a power law is bounded at high frequencies by flapping, which occurs at a frequency of about 2.8\,Hz. 
What bounds the scaling interval at low frequencies is not known, but the bound compares favorably with the frequency at which eagles traverse typical atmospheric eddies, which is about 0.1\,Hz given a 100\,m eddy and 10\,m/s flight speed. As described in the Materials \& Methods section, removing those data associated with flapping produced no change in the slope of the spectra.

The eagle’s acceleration spectra exhibit a logarithmic slope that is close to ‑5/3rds, 
a slope that does not meaningfully change when the spectra are conditioned on local wind speeds or on migration versus residency. 
Power law fits to the eagle’s acceleration spectra of the form, 
\begin{equation}
{\check{E}}_a\left(f\right)=Cf^n, 
\label{eqn:spect}
\end{equation}
where $f$ is the frequency, 
yield exponents $n = -1.59 \pm 0.07$ that are close to but slightly shallower than the value of $-5/3 \approx -1.67$ that is approached in 
the ``inertial range'' of 
turbulence toward higher intensities 
(\textit{i.e.} higher Reynolds numbers) \cite{Mydlarski1996,Pope2000}. 

We explain the acceleration spectra of the eagle by combining Newton’s 
Second Law 
of proportion between accelerations and forces with an assumption that changes in the aerodynamic forces on the eagle were approximately linear in changes of the relative velocity between the eagle and the air. 
That is, we suppose that velocity fluctuations, $w$, about the mean wind speed produce changes in the aerodynamics forces, $F_w \sim w$, that are linear in the velocity fluctuations at the leading order. 
This assumption holds under several conditions. 
For instance, the lift generated by a wing is linear in vertical perturbations to the wind vector until stall \cite{Hull2007}, the thrust produced by a propeller is linear in changes to its airspeed that are small relative to its self-induced wind \cite{Seddon2011}, and nonlinear drag exhibited at high Reynolds numbers is linear in small changes to airspeed. 
The extent to which these conditions apply to the eagle is uncertain 
and requires further investigation, 
but our findings support the view that the linear assumption is broadly useful 
despite the complexity of the interactions between the eagle and unsteady flow. 
Combined with Newton’s 
Second 
Law, $a_b = F_w/m_b$, 
where $m_b$ is the mass of the bird, 
it follows that the fluctuations in the accelerations of the eagle, 
\begin{equation}
a_b = w/\tau_b, \label{eqn:acc}
\end{equation}
have the same spectrum as the wind velocity fluctuations
, $w$, 
encountered by the eagle.  
Here, $\tau_b$ is the characteristic response time of the eagle to velocity fluctuations, which is larger for heavier or more aerodynamic eagles.
We also assume that the winds traversed by the eagle were an unbiased sample of all winds, 
and so exhibited a -5/3rds power law

The data show (Fig.~\ref{fig:spectra}) that higher wind speeds are associated with larger accelerations of the eagle within the interval of 0.2 to 2\,Hz, where scaling prevails in the spectra. 
We found that the fitted values for the coefficients, $C$, in Eq.~\ref{eqn:spect} increased monotonically with wind speed $W$ (Fig.~\ref{fig:spectra}, {\it Inset}). To perform the fits, we fixed the exponent, $n$, for the spectrum to -5/3, which tends to reduce scatter in $C$ relative to the fit with a free exponent we reported above. The acceleration spectrum does not vanish at zero wind speed, which may be due to turbulence produced by thermals, that arise even in the absence of a mean wind. 

We explain the larger accelerations in higher winds 
by the increasing strength of turbulence in these winds
(see Materials \& Methods). 
The key point is that the spectrum of the wind velocity is proportional to the 2/3rds power of the turbulent dissipation rate, 
which is itself proportional to the cube of the turbulence intensity, 
according to classical phenomenology \cite{Pope2000}. 
At a given altitude within the atmospheric boundary layer where the eagle flies, 
the turbulence intensity changes in proportion to wind speed, $W$. 
Taken all together with the linear relationship 
between eagle accelerations and wind speeds, 
it follows that the prefactor for the acceleration spectrum 
increases quadratically with wind speed: $(W^3)^{2/3} = W^2$. 
Given the eagle’s acceleration spectrum in the form given by Eq.~\ref{eqn:spect} 
with slope $n=-5/3$, 
the prefactor including all dimensional parameters is 
\begin{equation}
C\equiv\left(IW/\tau_b\right)^2(V/L)^{2/3} \label{eqn:C}
\end{equation}
where $I = w^\prime/W$ is the turbulence intensity, 
$w^\prime$ is the characteristic velocity of the turbulence, 
$L$ is the large-eddy length scale, 
$V$ is the airspeed, 
$C$ has units of $\mathrm{m^2/s^{14/3}}$, 
and we neglect dimensionless constants of order one 
including the Kolmogorov constant \cite{Sreenivasan1995}. 

We fit quadratic functions 
of the form given by Eq.~\ref{eqn:C}
to the migration and residency spectra separately and observe a qualitative difference between them 
(Fig.~\ref{fig:spectra}, inset). 
That is, the spectra were less responsive to increases in wind speed while the eagle was migrating than when it was in residence. 
The differences between these two modes of flight may be due to choices made by the eagle. 
For instance, during migration, birds tend to be more conservative about how they fly, 
tending to flap less to conserve energy. 

The eagle was presumably able to act at the full range of frequencies over which the accelerations exhibited a power law spectrum. We approximate the characteristic response time of the bird to velocity fluctuations, $\tau_b$, from the eagle’s acceleration spectra 
using Eq.~\ref{eqn:C}.  
Assuming typical turbulence intensities in the boundary layer of $I\approx10$\%, typical airspeeds of  $V \approx 10$\,m/s, large-eddy length scales of $L \approx 100$\,m, and using the quadratic fits shown in Fig.~\ref{fig:spectra}, we found that $\tau_b \approx$ 0.7\,s and 1.4\,s for non-migrating and migrating flight, respectively. These values for $\tau_b$ correspond approximately to the right boundary of the scaling interval in the spectra (Fig.~\ref{fig:spectra}) and to the timescale of flapping (0.4\,s).

\section*{Discussion}

To characterize the eagle's responsiveness to turbulence, 
we compare the eagle's response time estimated above, $\tau_b \approx 1$\,s, 
to timescales of the turbulence. 
The ratio forms an analog to the Stokes number for inertial particles, 
which is larger than one for heavy particles weakly affected by turbulence, 
such as cannonballs, 
and smaller than one for lightweight ones at the mercy of the wind, 
such as feathers \cite{bewley2013}. 
The Stokes number we estimate for the eagle 
is comparable to the one for the weakly inertial particles in Fig.~\ref{fig:PDF}, 
which suggests that the eagle is more a feather than a cannonball 
in the face of turbulent eddies larger than its wingspan. 
For eddies of the same size as the eagle ($l$ $\approx$ 2\,m), 
the turnover times, $\tau_l$ 
$\sim (L/w^\prime)(l/L)^{2/3}$ 
are about ten times smaller than they are for largest eddies \cite{Pope2000}, 
and the corresponding eagle Stokes number, $\tau_b / \tau_l$, 
is in the range of 0.01 to 0.1. 
Eddies smaller than the eagle 
do not act uniformly over the surface of the eagle 
so that estimating their effects requires more detailed considerations. 
In contrast to the Stokes number for inertial particles, 
which has an unambiguous origin in their governing equations, 
the utility of a Stokes number in describing eagle dynamics 
needs to be explored in future work.


We find that the eagle’s vertical displacements were roughly as large as those of the motions of the turbulent air, and that in the time during which it went up (or down), it flew forward about as far. 
Those displacements, $z$, are caused by accelerations associated with turbulence scale as $\tau^2a_b$ where the timescale, $\tau$, resides in the interval where turbulence scaling prevails in the spectrum and $a_b$ is a suitably filtered acceleration. For comparison, the air itself moved in a typical eddy by $\mathrm{\Delta}\sim\tau w$, so that the eagle’s displacement relative to this convective motion was $z/\mathrm{\Delta}\sim\tau a_b/w$. For linear responses to wind fluctuations (Eq.~\ref{eqn:acc}) this simplifies to $z/\mathrm{\Delta}\sim \tau/\tau_b$, which we can estimate since $\tau =1/f$ is between about ½ and 10\,s and since $\tau_b\approx 1$\,s, as estimated above. 
The ratio, $z/\mathrm{\Delta}$, is in the vicinity of one, and its difference from one depends on numerical prefactors that need to be determined by future experiments. 
Similar arguments comparing the displacements of the eagle to its mean motion, $V\tau$, lead to a similar conclusion: that the ratio scales with $\tau/\tau_b$, though in this case modified by the ratio between wind velocities and airspeed. These arguments suggest that the eagle’s interactions with air movements are substantial enough to make the difference between reaching or missing a target such as prey, for instance. 

Despite the complexity of an eagle’s behavior and its large mass, size, and speed relative to the turbulence around it, the eagle’s acceleration statistics can be explained primarily by its interactions with the turbulent air over a broad range of behaviorally relevant frequencies. 
This held even at low wind speeds when eagles often exploit thermal updrafts rather than the orographic updrafts eagles seek in higher winds. 
For high frequency accelerations, turbulent fluctuations are not uniform over the surface area of the eagle and are partially filtered by the inertia of the eagle. 
At low frequencies below 0.1~Hz, the eagle’s decisions about where to fly dominate its interactions with turbulence, and the turbulence itself loses its approximately universal statistical structure. 
At intermediate frequencies, eagles likely sense and respond to turbulence, and the evidence supports a proportional and linear response to wind velocity fluctuations that imprint themselves as a -5/3rds law in the spectrum of accelerations. 
Even for nonlinear (\textit{e.g.} quadratic) relationships, 
turbulence spectra exhibit -5/3rds power laws 
with larger prefactors at higher wind speeds \cite{VanAtta1975}, 
further supporting the dominance of turbulence among the forces on the eagle. 
The data suggest that rather than seeing turbulence as a disturbance inimical to the eagle’s path, 
it might be seen as a source of energy that the eagle skillfully knits into the act of flight and travel. 

The signature of turbulence was woven into every aspect of the eagle’s flight trajectory, 
and not removed, avoided, or filtered out. 
Our findings point to the importance of the instantaneous wind in the vicinity of the bird 
and of the bird’s characteristic response time to changes in the wind. 
Ultimately, our research reinforces the need to fully incorporate 
an understanding of turbulence into our understanding of eagle movements and behaviors, 
with implications for other flying and  swimming creatures. 
Volant animals smaller than eagles, for instance, 
are likely even more influenced by the energy in turbulence \cite{Ravi2015,Ristroph2010} 
and by the internal work of the wind \cite{Langley1893}. 
The details of how flying animals of all sizes have adapted to this energy source 
and how the energy exchanges with the wind compete with and interact 
with the energetics of flight maneuvers 
are some of the most interesting questions in bio-engineering. 
Part of understanding these organismal responses to environmental challenges at diverse scales can only come by the development of sensors that gather fine-scale data on both the movements of birds and of the air surrounding them. This development could see the deployment of environmental sensors, both on man-made vehicles and free-flying animals, that will revolutionize our understanding of environmental variation, especially within the aerosphere and the turbulent winds that distinguish it \cite{Edwards2016,Weimerskirch2020}.

\section*{Materials \& Methods}
\subsection*{Characteristics of the eagle}

We captured an adult female golden eagle ({\it Aquila chrysaetos}) using a NetLauncher (Trapping Innovations, LLC, Jackson, WY, USA) set up on a bait pile of road-killed dear near Heflin, Alabama, USA in January 2016. We banded the eagle with an aluminum U.S. Geological Survey bird band and took standard morphometric measurements. 
The eagle weighed 
5.2
\,kg with a wing chord of  
600
\,mm and a tail length of 
330
\,mm. 
Use of golden eagles for this research was approved by the West Virginia University Institutional Animal Care and Use Committee (protocol number 11-0304).

\subsection*{Acquisition of accelerations and positions}

We attached a CTT-1100 (Cellular Tracking Technologies, LLC, Rio Grande, NJ, USA) solar GPS (Global Positioning System) telemetry unit equipped with an accelerometer using a backpack style of attachment with Teflon ribbon (Bally Ribbon Mills, Bally, PA, USA). The unit weighed 70\,g. 
The unit collected a GPS position, which included location, ground speed, course over ground, altitude, and horizontal and vertical dilution of precision, every 15 minutes while the bird was stationary and every 6-7\,s while the bird was moving. From the 15$^{th}$ to the 31$^{st}$ of March 2016, the unit was programmed to collect acceleration data on the $x$, $y$, and $z$ axes at a rate of 40\,Hz (40 times per second). GPS coordinates were recorded at frequencies of up to 0.0167\,Hz. Due to limitations of the onboard storage, bursts of data were collected at irregular time intervals and resulted in approximately 17 hours of acceleration and GPS data during approximately 15 hours of flight. The data were stored on the unit and transmitted daily via the Global System for Mobile Communications (GSM) to CTT servers. We evaluated approximately 200 hours of GPS data. Over the 17-day study period, the eagle was in flight for 49.7\% (± 15.8\%) of each day. 

\subsection*{Determination of local wind speeds}

We obtained predictions for wind speeds encountered by the eagle using data from the National Centers for Environmental Prediction North American Regional Reanalysis (NCEP NARR). 
Wind speeds were measured by NCEP NARR every 3 hours at a height of 10\,m and at stations approximately 32\,km apart. 
For each set of GPS coordinates from the eagle, 
we used bilinear interpolation to estimate the wind speeds at the eagle’s position. 
We then linearly interpolated the conditions in time to estimate the wind speeds experienced by the eagle at the timestamps associated with each given set of GPS coordinates. 
Note that wind speeds generally increase with increasing altitude and we did not consider the altitude of the bird in this study. 
Flight altitude above mean sea level was normally distributed with a mean of 983.9 $\pm$ 345.1\,m. 
Mean flight altitude Above Ground Level (AGL) was 459.6 $\pm$ 328.0\,m. 
Higher flight altitudes tend to be associated with thermal soaring and gliding whereas lower flight altitudes are associated with slope soaring. 
The consequences of these altitude variations will be the subject of future studies. 

Since the acceleration and GPS data were recorded at different frequencies, 
we mapped the wind speeds to the acceleration data by linearly interpolating the wind speeds associated with the GPS coordinates in time. 
The time interval between the nearest GPS and acceleration data was at most one hour and typically about three minutes. 
The eagle experienced wind speeds ranging from 1.3 to 13.5\,m/s. 

\subsection*{Classification of flapping and non-flying behaviors}

We used characteristics of the acceleration data to determine the behavior of the bird. 
Using previous studies that analyzed patterns in accelerometry data \cite{Nathan2012,Ropert-Coudert2004,Sur2017}, we developed our own methods to classify flapping flight and non-flying behaviors such as taking-off, landing, 
or perching. 

Flapping flight is associated with large oscillations at the flapping frequency in both the $x$- and $z$-accelerations \cite{Laich2008,Ropert-Coudert2004}. As the eagle flaps its wings, the flapping motion results in the body of the bird oscillating up and down as well as forward and back due to the generation of thrust \cite{Pennycuick1975}. To identify flapping flight, first we approximated the flapping frequency using a Fast Fourier Transform on the $z$-accelerations and identified a large spike in the signal at 2.8\,Hz to be the average flapping frequency \cite{Halsey2009} with a harmonic seen at 5.2\,Hz. We then took all $x$-accelerations above -0.2\,g and fit a sinewave with a frequency equal to the average flapping frequency to the $z$-accelerations and surrounding points within 1⁄2 of the flapping period. If the correlation between the $z$-accelerations and the sine wave was above 0.6 then we labeled the accelerations, as well as all data within 3 half-periods, as flapping. 
For different combinations of the parameters listed above, only the spikes in the spectra were removed and the power law was not 
affected
, and we used the set that resulted in the removal of the spike at 2.8\,Hz but retained the most data in the non-flapping flight data set. Using these values, we found that the bird flapped 9.2\% of the time it was flying. 

Since we are interested in the interactions between birds and atmospheric turbulence, we identified and removed accelerations associated with non-flying behaviors such as take-off, landing, 
and perching. 
We hypothesized that these non-flying behaviors are associated with large pitch values, 
or rotations of the acceleration of gravity in the frame of the telemetry unit on the eagle, 
which was consistent with the data. 
While taking off and landing, an eagle rotates its body from the vertical toward the horizontal and vice versa. 
While perching, 
the body of the eagle is oriented toward the vertical. 
Therefore, to identify these nonflying behaviors, we first estimate the pitch of the eagle then label any pitch values above some threshold as nonflying. 

The tri-axial acceleration data includes both the static accelerations due to gravity and dynamic accelerations. We approximated the orientation of the static acceleration by smoothing the acceleration signal \cite{Wilson2008}. In the absence of other disturbances, only gravitational forces are measured; this allowed us to determine the orientation of the bird. To separate the static and dynamic accelerations, we used a running mean of 5\,s. Assuming that the bank angle is small, we estimated the pitch of the bird by taking the arcsine of the smoothed $x$-accelerations.
Note that the smoothed acceleration should not exceed one; when the running mean is too short, the smoothed accelerations may be too large. We found that the length of our running mean had little effect on our estimates for pitch and chose a length of 5\,s since it was the smallest running mean that produced values less than one. 

We then labeled accelerations associated with pitch angles larger than $25^{\circ}$ as non-flying behaviors. We found that, regardless of our choice for pitch limits, the spectra are generally unchanged, and the spectral slope is approximately constant. Removal of non-flying behaviors greatly decreased the value of the spectra for frequencies below 0.2\,Hz, which indicates that non-flying behaviors have a large influence on the energy at low frequencies, which we expect given that changes in body orientation are typically associated with lower frequencies \cite{Wilson2008}.

\subsection*{Estimation of the spectral prefactor}

To derive Eq.~\ref{eqn:C} in the main text, 
we begin with the wind velocity spectrum in the inertial range of turbulence, 
given at high Reynolds numbers by 
$E_w(\kappa) = C_k \epsilon^{2/3} \kappa^{-5/3}$, 
according to Kolmogorov's theory, 
where 
$\epsilon = C_\epsilon w^{\prime 3}/L$ is the dissipation rate 
of turbulence kinetic energy per unit mass, 
$\kappa$ is the wavenumber, 
$C_\epsilon$ and $C_\kappa$ are constants of order one, 
and $w^\prime$ and $L$ are characteristic velocity and length scales of turbulence, 
respectively \cite{Pope2000}. 
We write the turbulence intensity as $I=w^\prime/W$ 
so that the spectrum reads 
$E_w(\kappa) = C_k \left(C_\epsilon/L\right)^{2/3} \left( IW \right)^2 \kappa^{-5/3}$. %
According to Taylor’s hypothesis \cite{Taylor1938}, 
the frequency and wavenumber spectra are 
proportional 
and the eagle intercepted spatial structures of wavenumber $\kappa$ 
at frequencies $f = \kappa V$, 
where $V$ is the eagle’s mean airspeed. 
This relation holds approximately when 
$w^\prime \ll V$ \cite{Lumley1965}
. 
According to Eq.~\ref{eqn:acc} 
the spectra of the eagle’s accelerations 
and of the wind velocity fluctuations are proportional, 
so that
$\check{E}_a(f) = \check{E}_w(f)/\tau_b^2$. 
In this way, we find that 
\begin{equation}
{\check{E}}_a\left(f \right) 
    = C_k \left( C_\epsilon V/L \right)^{2/3} \left(IW/\tau_b\right)^2  f^{-5/3}. 
\end{equation}
Neglecting the order-one constants $C_\kappa$ and $C_\epsilon$ 
we observe that if $\check{E}_a(f) = C f^{-5/3}$, 
then the prefactor $C$ is as given in Eq.~\ref{eqn:C}. 

\subsection*{Data Availability}
The data that support the findings of this study are available from Cellular Tracking Technologies but restrictions apply to the availability of these data, which were used under license for the current study, and so are not publicly available. Data are however available from the authors upon reasonable request and with permission of Cellular Tracking Technologies and Friends of Talladega National Forest.

\subsection*{Acknowledgments}
We thank our colleagues for discussions, including K. Atell, I. Cohen, B. Mehlig, T. Schneider, J. Wang, and Z. Warhaft as well as J. Stober of the US Forest Service, Shoal Creek Ranger District, and E. Soehren and C. Threadgill of the Alabama Dept. of Conservation and Natural Resources. 
Funding for the capture and tagging of the eagle as well as for the purchase of the telemetry unit was provided by the Friends of Talladega National Forest. 

\medskip

\printbibliography

\end{document}